\documentclass{ws-procs975x65v2}

\def\ti#1{}

\def\etal {{\it et al.}}

\newcommand{\beq}{\begin{equation}}
\newcommand{\eeq}{\end{equation}}
\newcommand{\bea}{\begin{eqnarray}}
\newcommand{\eea}{\end{eqnarray}}

\begin{document}

\title{The Standard-Model Extension and Tests of Relativity
\footnote{Proceedings of 11th Marcel Grossmann Meeting,
Berlin, Germany, 23-29 July, 2006.}}

\author{NEIL RUSSELL
}

\address{Physics Department,
Northern Michigan University,
\\
Marquette, MI 49855, USA
\\
nrussell@nmu.edu}

\begin{abstract}
The Standard-Model Extension, or SME,
is a general framework for the study of Lorentz violation in physics.
A broad variety of experiments is able to access the SME coefficient space.
This proceedings briefly summarizes theory and experiments
aimed at testing Special Relativity by measuring these coefficients.
\end{abstract}


\bodymatter

Lorentz symmetry is a central feature of the existing
theories of gravitation and particle physics.
The existence of highly sensitive experiments
with the ability to test Lorentz symmetry at
unprecedented levels
raises the possibility of discovering
unconventional effects.
This is clearly of interest to physicists since
it may pave the way to finding a unified theory of quantum gravity.

A series of publications since 1989
has established a framework,
the Standard-Model Extension,
or SME,
that provides a detailed description
of possible Lorentz violations in nature
in the context of effective field theory.
At the basic level,
this work focuses on
a variety of theoretical issues,
including string theory,
and spontaneous symmetry breaking.\cite{th1}
Much theoretical and experimental effort has
been directed towards the study of Lorentz symmetry
in Minkowski space,
for which the effective field theory is an
extension of the Standard Model of particle physics.
In flat spacetime,
the SME comprises a broad variety of constant coefficients
for Lorentz violation that can in principle be measured.\cite{SME:minkowski}
These coefficients transform as conventional Lorentz tensors
under observer transformations, but under rotations and boosts
of experimental systems, called particle transformations,
they are not transformed.

An important category of experimental symmetry tests involves
searching for couplings between the electron spin
and the Lorentz-violating SME background.
The basic idea is that the radiation released in a
transition between different spin states
has frequency that depends on the spin quantization axis
and that differs for particles and antiparticles.
Consequently, spectral transitions in atoms with controlled quantization axes,
such as occur in atomic clocks,
are well suited to tests of Lorentz symmetry.
To see small variations in the output frequency of
a sensitive clock,
one has to compare it to the output of another clock
for which the effects are absent, or at least different.
So, such experiments are often called
clock-comparison experiments.\cite{cc}
One of the common scenarios involves
monitoring the outputs for long enough to detect
the sidereal effects associated with the rotation of the apparatus
relative to the distant stars.
Tests and theoretical investigations based on these ideas
include ones done for
hydrogen masers,
antihydrogen,
noble-gas masers,
space-mounted atomic clocks,
Penning traps,
and
torsion pendula.\cite{electr2}

The effects of Lorentz-violation on the electromagnetic sector
are described by 19 coefficients at leading order and
are amenable to sensitive experimental investigations.
Analysis of birefringence data from cosmological sources
has placed stringent limits on 10 of these,
while optical and microwave cavity resonators
have placed limits on the remaining ones.\cite{photon:cavities}
Cosmological birefringence tests are based on
distant processes producing the radiation,
but offer fantastic sensitivities.
Laboratory cavity experiments have undergone numerous
innovations to improve their experimental reach,
including cryogenic cooling,
the use of optical sapphire crystals,
and placement on rotating turntables to exploit geometrical properties.
Other investigations involving photons include,
for example, ones based on
Cerenkov radiation,
synchrotron radiation,
Compton scattering,
and Doppler-shift experiments.\cite{photon2}

Lorentz symmetry has also been tested
in the context of various other particles.
For example,
in the case of neutrinos,
simple models constructed from
the SME coefficients have been found to
be consistent with known neutrino data
while offering the advantage of fewer parameters
and masses.\cite{neutrino}
Accelerator-related physics investigations
of Lorentz symmetry include ones with
a variety of neutral mesons
and others with muons.\cite{mesons_muons}
Further details of Lorentz tests
in flat spacetime can be found in various
overview sources.\cite{SMEreviews}

The gravitational sector of the Standard-Model Extension
consists of a framework for addressing Lorentz and CPT violation
in curved spacetimes,
including ones with torsion.\cite{gr}
The coefficients for Lorentz violation
typically vary with position,
adding complexity to the manner
in which matter couples to the background. %
To set up the framework for the full
Standard-Model Extension,
the vierbein formalism can be adopted,
since it allows the spinor properties of ordinary matter
to be incorporated.
It also has the useful feature of distinguishing
naturally between local Lorentz transformations
and general coordinate transformations.
Lorentz symmetry breaking must be either explicit or spontaneous.
A study of this topic has shown that explicit Lorentz violation,
in which the breaking occurs in the Lagrangian density,
is incompatible with generic Riemann-Cartan spacetimes.
On the other hand, spontaneous breaking
can be successfully introduced in a consistent manner.
One of the far-reaching results associated with
spontaneous Lorentz breaking is that
it always goes hand in hand with spontaneous breaking
of diffeomorphism symmetry.
The 10 possible Nambu-Goldstone modes
associated with the
six generators for Lorentz transformations and
the four generators for diffeomorphisms
have been studied.
The fate of these modes depends on the spacetime geometry
and the dynamics of the tensor field triggering the spontaneous Lorentz violation.
The results are consistent with the known massless particles in nature,
the photon and the graviton.
An extensive study has been made of the pure-gravity sector of the
SME with the aim of finding possible experimental consequences.
Of particular interest are experiments involving
lunar and satellite laser ranging,
laboratory tests with gravimeters and torsion pendula,
measurements of the spin precession of orbiting gyroscopes,
timing studies of signals from binary pulsars,
and the classic tests
involving the perihelion precession
and the time delay of light.
The sensitivity range of these experiments
is parts in $10^4$ to parts in $10^{15}$.


\begin{thebibliography}{00}
\bibitem{th1}
  V.A.\ Kosteleck\'y and R.\ Potting,
   \ti{``CPT, strings, and meson factories,''}
  Phys.\ Rev.\ D {\bf 51}, 3923 (1995);
%
  \ti{``Expectation Values, Lorentz Invariance, and CPT in the Open Bosonic
  String,''}
  Phys.\ Lett.\ B {\bf 381}, 89 (1996);
  \ti{``Analytical construction of a nonperturbative vacuum for the open bosonic
   string,''}
  Phys.\ Rev.\ D {\bf 63}, 046007 (2001);
%
   \ti{``CPT and strings,''}
  Nucl.\ Phys.\ B {\bf 359}, 545 (1991);
%
  V.A.\ Kosteleck\'y and S.\ Samuel,
   \ti{``Spontaneous Breaking Of Lorentz Symmetry In String Theory,''}
  Phys.\ Rev.\ D {\bf 39}, 683 (1989);
%
   \ti{``Phenomenological Gravitational Constraints On Strings And Higher
   Dimensional Theories,''}
  Phys.\ Rev.\ Lett.\  {\bf 63}, 224 (1989);
%
   \ti{``Gravitational Phenomenology In Higher Dimensional Theories And Strings,''}
  Phys.\ Rev.\ D {\bf 40}, 1886 (1989);
%
  \ti{``Photon And Graviton Masses In String Theories,''}
  Phys.\ Rev.\ Lett.\  {\bf 66}, 1811 (1991);
%
 B.\ Altschul and V.A.\ Kosteleck\'y,
 \ti{Spontaneous Lorentz Violation and Nonpolynomial Interactions,}
 Phys.\ Lett.\ B {\bf 628}, 106 (2005).

\bibitem{SME:minkowski}
  D.\ Colladay and V.A.\ Kosteleck\'y,
   \ti{``CPT violation and the standard model,''}
  Phys.\ Rev.\ D {\bf 55}, 6760 (1997);
%
   \ti{``Lorentz-violating extension of the standard model,''}
  Phys.\ Rev.\ D {\bf 58}, 116002 (1998);
%
  V.A.\ Kosteleck\'y and R.\ Lehnert,
  \ti{``Stability, causality, and Lorentz and CPT violation,''}
  Phys.\ Rev.\ D {\bf 63}, 065008 (2001).


\bibitem{cc}
  V.A.\ Kosteleck\'y and C.D.\ Lane,
   \ti{``Constraints on Lorentz violation from clock-comparison experiments,''}
  Phys.\ Rev.\ D {\bf 60}, 116010 (1999);
%
%
  P.~Wolf \etal,
  \ti{``Cold atom clock test of Lorentz invariance in the matter sector,''}
  Phys.\ Rev.\ Lett.\  {\bf 96}, 060801 (2006);
%
  F.\ Cane {\it et al.},
  \ti{``Bound on Lorentz- and CPT-Violating Boost Effects for the Neutron,''}
  Phys.\ Rev.\ Lett.\  {\bf 93}, 230801 (2004);
%
  D.F.\ Phillips \etal,
  \ti{``Limit on Lorentz and CPT violation of the proton using a hydrogen  maser,''}
  Phys.\ Rev.\ D {\bf 63}, 111101 (2001);
%
  M.A.\ Humphrey \etal,
  \ti{``Testing Lorentz and CPT symmetry with hydrogen masers,''}
  Phys.\ Rev.\ {\bf A 68}, 063807 (2003);
%
  \ti{``Double-resonance frequency shift in a hydrogen maser,''}
  Phys.\ Rev.\  {\bf A62}, 063405 (2000);
%
  D.\ Bear \etal,
  \ti{``Limit on Lorentz and CPT violation of the neutron using a two-species
   noble-gas maser,''}
  Phys.\ Rev.\ Lett.\  {\bf 85}, 5038 (2000).
%

\bibitem{electr2}
  G.M.\ Shore,
   \ti{``Strong equivalence, Lorentz and CPT violation, anti-hydrogen  spectroscopy
   and gamma-ray burst polarimetry,''}
  Nucl.\ Phys.\ B {\bf 717}, 86 (2005);
%
  R.\ Bluhm \etal,
   \ti{``CPT and Lorentz tests in hydrogen and antihydrogen,''}
  Phys.\ Rev.\ Lett.\  {\bf 82}, 2254 (1999);
%
   \ti{``Clock-comparison tests of Lorentz and CPT symmetry in space,''}
  Phys.\ Rev.\ Lett.\  {\bf 88}, 090801 (2002);
%
   \ti{``Probing Lorentz and CPT violation with space-based experiments,''}
  Phys.\ Rev.\ D {\bf 68}, 125008 (2003);
%
  \ti{ ``Testing CPT with anomalous magnetic moments,''}
  Phys.\ Rev.\ Lett.\  {\bf 79}, 1432 (1997);
%
  \ti{ ``CPT and Lorentz tests in Penning traps,''}
  Phys.\ Rev.\ D {\bf 57}, 3932 (1998);
%
%
H.\ Dehmelt \etal,
   \ti{``Past electron positron g-2 experiments yielded sharpest bound on CPT
    violation,''}
  Phys.\ Rev.\ Lett.\  {\bf 83}, 4694 (1999);
%
R.K.\ Mittleman \etal,
  \ti{``Bound on CPT and Lorentz symmetry with a trapped electron,''}
  Phys.\ Rev.\ Lett.\  {\bf 83}, 2116 (1999);
%
G.\ Gabrielse \etal,
   \ti{``Precision mass spectroscopy of the antiproton and proton using
    simultaneously trapped particles,''}
  Phys.\ Rev.\ Lett.\  {\bf 82}, 3198 (1999);
%
%
R.\ Bluhm and V.A.\ Kosteleck\'y,
  \ti{ ``Lorentz and CPT tests with spin-polarized solids,''}
  Phys.\ Rev.\ Lett.\  {\bf 84}, 1381 (2000);
%
B.R.\ Heckel \etal,
  \ti{``New CP-violation and preferred-frame tests
  with polarized electrons,''}
  Phys.\ Rev.\ Lett.\  {\bf 97}, 021603 (2006);
%
 L.-S.\ Hou \etal,
 \ti{``Test of Cosmic Spatial Isotropy for Polarized Electrons Using
 a Rotatable Torsion Balance''}
 Phys.\ Rev.\ Lett. {\bf 90}, 201101 (2003);
%
D.\ Colladay and P.\ McDonald,
  \ti{``Bose-Einstein condensates as a probe for Lorentz violation,''}
  Phys.\ Rev.\ D {\bf 73}, 105006 (2006).

\bibitem{photon:cavities}
%
S.\ Herrmann \etal,
  \ti{``Test of the isotropy of the speed of light using a continuously rotating
  optical resonator,''}
  Phys.\ Rev.\ Lett.\  {\bf 95}, 150401 (2005);
%
%
  P.L.\ Stanwix \etal,
  \ti{``Test of Lorentz invariance in electrodynamics using rotating cryogenic
  sapphire microwave oscillators,''}
  Phys.\ Rev.\ Lett.\  {\bf 95}, 040404 (2005);
%
  P.\ Antonini \etal,
  \ti{``Test of constancy of speed of light with rotating cryogenic optical
   resonators,''}
  Phys.\ Rev.\ A {\bf 71}, 050101 (2005);
%
  \ti{``Reply to ``Comment on `Test of constancy of speed of light with rotating
  cryogenic optical resonators',''}
  Phys.\ Rev.\ A {\bf 72}, 066102 (2005);
%
  M.E.\ Tobar \etal,
  \ti{``New methods of testing Lorentz violation in electrodynamics,''}
  Phys.\ Rev.\ D {\bf 71}, 025004 (2005);
%
%
  \ti{``Comment on Test of constancy of speed of light with rotating cryogenic
  optical resonators,''}
  Phys.\ Rev.\ A {\bf 72}, 066101 (2005);
%
  P.\ Wolf \etal,
   \ti{``Improved test of Lorentz invariance in electrodynamics,''}
  Phys.\ Rev.\ D {\bf 70}, 051902 (2004);
%
  \ti{``Whispering gallery resonators and tests of Lorentz invariance,''}
  Gen.\ Rel.\ Grav.\  {\bf 36}, 2352 (2004);
%
  H.\ M\"uller \etal,
  \ti{``Optical cavity tests of Lorentz invariance for the electron,''}
  Phys.\ Rev.\ D {\bf 68}, 116006 (2003);
%
 \ti{``Modern Michelson-Morley Experiment using Cryogenic Optical Resonators,''}
 Phys.\ Rev.\ Lett. {\bf 91} 020401 (2003);
%
  J.A.\ Lipa \etal,
  \ti{``A New Limit on Signals of Lorentz Violation in Electrodynamics,''}
  Phys.\ Rev.\ Lett.\  {\bf 90}, 060403 (2003);
%
  V.A.\ Kosteleck\'y and M.\ Mewes,
   \ti{``Cosmological constraints on Lorentz violation in electrodynamics,''}
  Phys.\ Rev.\ Lett.\  {\bf 87}, 251304 (2001);
%
   \ti{``Signals for Lorentz violation in electrodynamics,''}
  Phys.\ Rev.\ D {\bf 66}, 056005 (2002);
%
   \ti{``Sensitive polarimetric search for relativity violations in gamma-ray
  bursts,''}
  Phys.\ Rev.\ Lett.\ {\bf 97}, 140401 (2006).


\bibitem{photon2}
%
  R.\ Lehnert and R.\ Potting,
   \ti{``Vacuum Cerenkov radiation,''}
  Phys.\ Rev.\ Lett.\  {\bf 93}, 110402 (2004);
%
   \ti{``The Cerenkov effect in Lorentz-violating vacua,''}
  Phys.\ Rev.\ D {\bf 70}, 125010 (2004);
%
 B.\ Altschul,
  \ti{``Limits on Lorentz violation from synchrotron and
  inverse Compton sources,''}
  Phys.\ Rev.\ Lett.\ {\bf 96}, 201101 (2006);
%
  \ti{``Synchrotron and inverse Compton constraints on Lorentz violations for
  electrons,''}
  Phys.\ Rev.\ D {\bf 74}, 083003 (2006);
%
%
C.D.\ Lane,
   \ti{``Probing Lorentz violation with Doppler-shift experiments,''}
  Phys.\ Rev.\ D {\bf 72}, 016005 (2005).


\bibitem{neutrino}
  V.A.\ Kosteleck\'y and M.\ Mewes,
   \ti{``Lorentz and CPT violation in neutrinos,''}
  Phys.\ Rev.\ D {\bf 69}, 016005 (2004);
%
   \ti{``Lorentz and CPT violation in the neutrino sector,''}
  Phys.\ Rev.\ D {\bf 70}, 031902 (2004);
%
   \ti{``Lorentz violation and short-baseline neutrino experiments,''}
  Phys.\ Rev.\ D {\bf 70}, 076002 (2004);
%
  T.\ Katori \etal,
  \ti{``Global three-parameter model for neutrino oscillations using Lorentz
  violation,''}
  Phys.\ Rev.\ D {\bf 74}, 105009 (2006);
%
LSND Collab.,\  
   \ti{``Tests of Lorentz violation in anti-nu/mu $\to$ anti-nu/e oscillations,''}
  Phys.\ Rev.\ D {\bf 72}, 076004 (2005).
%

\bibitem{mesons_muons}
%
OPAL Collab.,\  
   \ti{``A study of B meson oscillations using hadronic Z0 decays containing
   leptons,''}
  Z.\ Phys.\ C {\bf 76}, 401 (1997);
%
%
BABAR Collab.,\ 
  \ti{``Search for CPT and Lorentz violation in B0 anti-B0 oscillations with
  inclusive dilepton events,''}
  hep-ex/0607103;
%
FOCUS Collab.,\ 
   \ti{``Charm system tests of CPT and Lorentz invariance with FOCUS. ((B)),''}
  Phys.\ Lett.\ B {\bf 556}, 7 (2003);
%
  V.A.\ Kosteleck\'y,
   \ti{``Sensitivity of CPT tests with neutral mesons,''}
  Phys.\ Rev.\ Lett.\  {\bf 80}, 1818 (1998);
%
   \ti{``Signals for CPT and Lorentz violation in neutral-meson oscillations,''}
  Phys.\ Rev.\ D {\bf 61}, 016002 (2000);
%
   \ti{``Formalism for CPT, T, and Lorentz violation in neutral-meson
   oscillations,''}
  Phys.\ Rev.\ D {\bf 64}, 076001 (2001);
%
  D.\ Colladay and V.A.\ Kosteleck\'y,
  \ti{``Tests of direct and indirect CPT violation at a B factory,''}
  Phys.\ Lett.\ B {\bf 344}, 259 (1995);
%
  \ti{``Testing CPT with the Neutral-D System,''}
  Phys.\ Rev.\ D {\bf 52}, 6224 (1995);
%
  V.A.\ Kosteleck\'y and R.\ Van Kooten,
  \ti{``Bounding CPT Violation in the Neutral-B System,''}
  Phys.\ Rev.\ D {\bf 54}, 5585 (1996);
%
N.\ Isgur \etal, 
  \ti{``Background enhancement of CPT reach at an asymmetric Phi factory,''}
  Phys.\ Lett.\ B {\bf 515}, 333 (2001);
%
  R.\ Bluhm, V.A.\ Kosteleck\'y and C.D.\ Lane,
   \ti{``CPT and Lorentz tests with muons,''}
  Phys.\ Rev.\ Lett.\  {\bf 84}, 1098 (2000);
%
  V.W.\ Hughes \etal,
   \ti{``Test of CPT and Lorentz invariance from muonium spectroscopy,''}
  Phys.\ Rev.\ Lett.\  {\bf 87}, 111804 (2001);
%
Muon g-2 Collab.,\ 
  \ti{``Testing CPT and Lorentz invariance with the anomalous
  spin precession of the muon,''}
  hep-ex/0110044.

\bibitem{SMEreviews}
  R.\ Bluhm,
   \ti{``Overview of the SME: Implications and phenomenology of Lorentz
   violation,''}
  arXiv:hep-ph/0506054;
%
%
D.\ Mattingly,
  \ti{``Modern tests of Lorentz invariance,''}
  Living Rev.\ Rel.\  {\bf 8}, 5 (2005).
%

%
\bibitem{gr}
  V.A.\ Kosteleck\'y,
  \ti{``Gravity, Lorentz violation, and the standard model,''}
  Phys.\ Rev.\ D {\bf 69}, 105009 (2004);
%
  R.\ Bluhm and V.A.\ Kosteleck\'y,
  \ti{``Spontaneous Lorentz violation, Nambu-Goldstone modes, and gravity,''}
  Phys.\ Rev.\ D {\bf 71}, 065008 (2005);
%
  V.A.\ Kosteleck\'y and R.\ Potting,
  \ti{``Gravity from local Lorentz violation,''}
  Gen.\ Rel.\ Grav.\ {\bf 37}, 1675 (2005);
%
  Q.G.\ Bailey and V.A.\ Kosteleck\'y,
  \ti{``Signals for Lorentz violation in post-Newtonian gravity,''}
  Phys.\ Rev.\ D {\bf 74}, 045001 (2006).

\end{thebibliography}
\end{document}